\documentclass[twocolumn,aps,pra,amsmath,amssymb,showpacs,superscriptaddress]{revtex4}

\usepackage{graphicx}
\usepackage{graphics}
\usepackage{latexsym}
\usepackage{amsmath}
\usepackage{amssymb}

\usepackage{longtable}
\usepackage{color}
\usepackage{amsmath}
\usepackage{amssymb}

\newcommand{\beq}{\begin{equation}}
\newcommand{\eeq}{\end{equation}}
\newcommand{\beqa}{\begin{eqnarray}}
\newcommand{\eeqa}{\end{eqnarray}}
\newcommand{\ket} [1] {\vert#1\rangle}

\def\ket#1{|#1\rangle}

\def\opone{\leavevmode\hbox{\small1\kern-3.8pt\normalsize1}}

\begin{document}
\title{Multiparty multilevel energy-time entanglement}

\author{Giuseppe Vallone}
\homepage{http://quantumoptics.phys.uniroma1.it/}
\affiliation{Centro Studi e Ricerche ``Enrico Fermi'', Via Panisperna 89/A, Compendio del Viminale, Roma 00184, Italy}
\affiliation{Dipartimento di Fisica, Universit\`{a} Sapienza di Roma, Piazzale Aldo Moro 5, I-00185 Roma, Italy}

\author{Paolo Mataloni}
\homepage{http://quantumoptics.phys.uniroma1.it/}
\affiliation{Dipartimento di Fisica, Universit\`{a} Sapienza di Roma, Piazzale Aldo Moro 5, I-00185 Roma, Italy}
\affiliation{Istituto Nazionale di Ottica Applicata (INOA-CNR), L.go
E. Fermi 6, 50125 Florence, Italy}

\author{Ad\'{a}n Cabello}
\email{adan@us.es}
\affiliation{Departamento de F\'{\i}sica Aplicada II, Universidad de Sevilla, E-41012 Sevilla, Spain}

\date{\today}

\begin{abstract}
Franson-like setups are inadequate for multiparty Bell
experiments with energy-time entanglement because postselected
events can depend on the local settings, and local models can
exploit this feature to reproduce the quantum predictions, even
in the case of ideal devices. We extend a previously introduced
interferometric scheme [A. Cabello {\em et al.}, Phys.\ Rev.\
Lett.\ {\bf 102}, 040401 (2009)] to solve this problem in the
$n$-qubit and $n$-qu$n$it cases. In addition, the proposed
setups allow us to prepare and test $n$-qubit
Greenberger-Horne-Zeilinger and $(\sum_{i=1}^n |i \ldots
i\rangle)/\sqrt{n}$ energy-time entangled states.
\end{abstract}


\maketitle


\section{Introduction}


Franson \cite{fran89prl} showed how the essential uncertainty
in the time of emission of a pair of particles can be exploited
to make undistinguishable two alternative paths that the
particles can take, and create what is called ``energy-time''
or ``time-bin'' \cite{bren99prl} entanglement, depending on the
method used to have uncertainty in the time of emission.
Franson proposed an experiment to demonstrate the violation of
the Bell Clauser-Horne-Shimony-Holt (CHSH) inequality
\cite{clau69prl} using energy-time entanglement. However, Aerts
{\em et al.} \cite{aert01prl} (see also \cite{cabe09prl})
showed that, even in the ideal case of perfect preparation and
perfect detection efficiency, there are local hidden variable
(HV) models that reproduce the quantum predictions for
Franson's test of the Bell-CHSH inequality. The reason is that,
in Franson's setup, the fact that photons are detected in
coincidence can depend on the local settings. This can be
exploited to build local HV models which simulate the quantum
predictions (see \cite{aert01prl, cabe09prl} for details).

Recently \cite{fran09pra}, Franson has argued that these local
HV models are not realistic in the sense of Einstein, Podolsky,
and Rosen (EPR) \cite{epr35pr}, because they do not describe
the path taken by the photons. However, in the Franson
Bell-CHSH experiment, the path taken by one photon cannot be
predicted with certainty from a measurement on the distant
photon, thus the path is not an element of reality in the sense
of EPR. The assumption that the local models must describe the
paths taken by the photons is an extra assumption which is not
necessarily satisfied by all local HV models. Actually, this
extra assumption is equivalent to the extra assumption that the
fact that a photon is detected at a specific time is
independent of the local experiment performed on that photon,
previously suggested in Ref.~\cite{aert01prl} as a way to avoid
the problem. The Franson Bell-CHSH experiment can rule out
local HV with this extra assumption, but cannot rule out local
HV models without this assumption.

Three different strategies have been proposed to solve this
problem:

Aerts {\em et al.} \cite{aert01prl} proved that local HV models
can be ruled out using a Franson's setup with a very fast local
switching if, instead of testing the standard two-setting
Bell-CHSH inequality, a specific three-setting Bell inequality
is tested. This solution has two problems: it requires a very
difficult to achieve fast switching, and also requires to
obtain experimentally a violation which is very close to the
maximum quantum violation obtained assuming ideal equipment
(i.e., it requires nearly perfect visibility). To our
knowledge, so far there no experimental implementation of this
proposal.

Brendel {\em et al.} \cite{bren99prl} proposed a modification
of Franson's setup which, in principle, solves the problem. The
modification consists of replacing the two beam splitters which
are closer to the source, by switchers synchronized with the
source. However, to our knowledge, these active switchers are
not available for photonic sources, thus in actual experiments
they are are replaced by passive beam splitters (see, e.g.,
\cite{bren99prl}), so the resulting setup suffers from the same
problem the original Franson's setup has. Recently, it has been
pointed out that active switchers could be feasible if photons
are replaced by molecules \cite{gnei08prl}.

More recently \cite{cabe09prl}, we have proposed a more radical
modification of Franson's setup which solves the problem and
can be actually implemented in the laboratory with photons. In
our scheme, both the short path of the first (second) particle
and the long path of the second (first) particle ends in
Alice's (Bob's) detectors. Then, the selection of events is
local (i.e., it does not require communication between Alice
and Bob), since coincidences occur every time the local
observer detects only one particle and, more important, is
independent of the local settings (see the details in Ref.
\cite{cabe09prl}). This last property is the one that makes
that this scheme do not suffer from the postselection loophole
that affect {\em all} previous Bell-CHSH experiments with
energy-time or time-bin entangled photons. This scheme has been
recently implemented in the laboratory \cite{lima09prl} and has
inspired a new source of electronic entanglement
\cite{frus09prb}.

The aim of this paper is to extend this scheme to the
multipartite case and discuss its applications for testing an
important class of multipartite Bell inequalities using
energy-time entanglement, and preparing some multipartite
multilevel energy-time entangled states relevant for quantum
information processing.

The paper is organized as follows. In Sec.~\ref{Sec1}, we show
that some previously proposed multipartite Franson-like
configurations are inadequate for testing multipartite Bell
inequalities of Mermin's type \cite{merm90prl} with energy-time
(and time-bin) entanglement. In Sec.~\ref{Sec2}, we introduce a
new scheme for creating Greenberger-Horne-Zeilinger (GHZ)
\cite{gree89pro} energy-time states and test Mermin's
tripartite Bell inequality without the problem previous
proposals have. In Sec.~\ref{Sec3}, we extend the scheme to
create three-qutrit energy-time entangled states, and then
generalize the setup to prepare $N$-qu$N$its energy-time
entangled states of the form $(\sum_{i=1}^n |i \ldots
i\rangle)/\sqrt{n}$. In Sec.~\ref{Sec4}, we discuss the sources
required for generating simultaneously $n$ particles with an
unknowable time of emission and discuss some problems appearing
when the $n>2$. Finally, in Sec.~\ref{Sec5}, we present our
conclusions.


\section{\label{Sec1}Franson-like configurations are inadequate for multipartite
Bell experiments}

\begin{figure}[h]
\begin{center}
\includegraphics[width=\columnwidth]{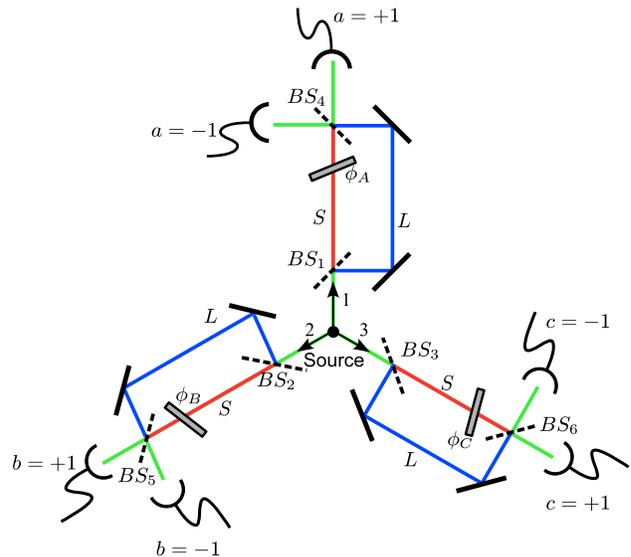}
\caption{Setup for a Franson-like energy-time three-party GHZ experiment.
The source emits three particles ($1$, $2$, and $3$) at the same unknown time.
Each of them is fed into an unbalanced interferometer with a short ($S$) and a long ($L$) path.
The essential uncertainty in the time of emission makes undistinguishable the case
where the three photons are detected at time $t_0$ after the time of emission ($SSS$) from the case where
the three photons are detected at time $t_1=t_0+\Delta t$ after the time of emission ($LLL$).}
\label{Fig1}
\end{center}
\end{figure}

Franson-like configurations for $n=3$ and $n=4$ particles have
been proposed in Refs.~\cite{shih93pla, pitt96pro}. The
simplest case, $n=3$, is illustrated in Fig.~\ref{Fig1}. In
these configurations, each of the $n$ parties is at the end of
an interferometer with a short path $S$ and a long path $L$,
and particle $i$ always ends in party $P_i$'s detectors.
Similar configurations can be easily constructed for $n \ge 3$
parties by adding more arms \cite{shih93pla, pitt96pro}. In
this Section we assume that we have a source emitting
simultaneously three particles at an unknown time (actual
sources with approximately this property will be discussed in
Sec.~\ref{Sec4}).


The setup in Fig.~\ref{Fig1} is inadequate for testing the
three-party Bell-Mermin inequality \cite{merm90prl} inspired by
GHZ proof of quantum nonlocality \cite{gree89pro}. The
three-party Bell-Mermin inequality is
\begin{equation}
\mu := |\langle A_0 B_0 C_1 \rangle + \langle A_0 B_1 C_0
\rangle + \langle A_1 B_0 C_0 \rangle - \langle A_1 B_1 C_1 \rangle|
\le 2, \label{Mermin}
\end{equation}
where $A_0$ and $A_1$ are dichotomic observables with possible
values $+1$ or $-1$ on Alice's qubit, $B_0$ and $B_1$ are
dichotomic observables on Bob's qubit, and $C_0$ and $C_1$ are
dichotomic observables on Carol's qubit.

According to quantum mechanics, the largest violation of
inequality \eqref{Mermin} is obtained by preparing the GHZ
state
 \beq
 \ket{{\rm GHZ}}=\frac{1}{\sqrt2}(\ket{SSS}+\ket{LLL}),
 \label{GHZ3}
 \eeq
which is an eigenstate with eigenvalue $-1$ of $\sigma_x^{(1)}
\otimes \sigma_y^{(2)} \otimes \sigma_y^{(3)}$, $\sigma_y^{(1)}
\otimes \sigma_x^{(2)} \otimes \sigma_y^{(3)}$, $\sigma_y^{(1)}
\otimes \sigma_y^{(2)} \otimes \sigma_x^{(3)}$, and
$-\sigma_x^{(1)} \otimes \sigma_x^{(2)} \otimes
\sigma_x^{(3)}$, and measuring the following six local
observables:
 \begin{subequations}
 \begin{align}
 &A_0=\sigma_y^{(1)}, &A_1=\sigma_x^{(1)},\\
 &B_0=\sigma_y^{(2)}, &B_1=\sigma_x^{(2)}, \\
 &C_0=\sigma_y^{(3)}, &C_1=\sigma_x^{(3)},
 \end{align}
 \end{subequations}
Then, quantum mechanics predicts $\mu =4$, which maximally
violates inequality \eqref{Mermin}.

The setup of Fig.~\ref{Fig1} can be used to produce the GHZ
state (\ref{GHZ3}) by postselecting threefold coincidences
(i.e., those events in which all three photons are detected at
the same time). This occurs in $25\%$ of the cases. In the
other cases, with equal frequencies, either two photons are
detected at time $t_0$ and one photon is detected at time
$t_1=t_0+\Delta t$, or one photon is detected at time $t_0$ and
two photons are detected at time $t_1$. The parties must store
the coincident events and reject the other events.

However, Table \ref{TableI} shows a local HV model which
reproduces the quantum predictions and, in particular, gives
$\mu =4$. In the model, for each local measurement, the
outcomes $S+$ (denoting that the photon will be detected at
time $t_0$ in the detector $+1$), $S-$, $L+$, and $L-$
(denoting that the photon will be detected at time $t_1$ in the
detector $-1$) are obtained with equal probability (as
predicted by quantum mechanics). $3/4$ of the events are
rejected during the postselection procedure because in that
cases not all three photons are detected. For the selected
events, $\mu=4$, which is the violation predicted by quantum
mechanics for an ideal experiment. Actually, similar local HV
models can be constructed to simulate any value $\mu \le 4$.
Moreover, similar local HV models can be constructed to
simulate any quantum prediction for any $n$-party Mermin
inequality using a Franson-like configuration like the one in
Fig.~\ref{Fig1}, but with an arbitrary number $n \ge 2$ of
parties, each of them with two settings.


\begin{table*}[htb]
\caption{\label{TableI}$1536$ sets of instructions of the local
HV model. Each row represents $96$ sets of local instructions
(first $6$ entries) and their corresponding contributions for
the calculation of $\mu$ after applying the postselection
procedure (last $4$ entries). In the first row, $L, L, L/S$
denotes the $48$ sets with three $L$ or two $L$ and one $S$,
with all possible combinations of signs: $L+, L+, L+; L+, L+,
S+;L+, S+, L+;\ldots;S-, L-, L-$. The other $48$ sets are
obtained by interchanging $S$ and $L$.}
\begin{ruledtabular}
{\begin{tabular}{ccccccccccc}
 $A_0$ & $A_1$ & $B_0$ & $B_1$ & $C_0$ & $C_1$
 & $\langle A_0 B_0 C_1 \rangle$ & $\langle A_0 B_1 C_0 \rangle$
 & $\langle A_1 B_0 C_0 \rangle$ & $\langle A_1 B_1 C_1 \rangle$ \\
\hline \hline
$S+$ & $L$ & $S+$ & $L$ & $L/S$ & $S+$ & $+1$ & rejected & rejected & rejected \\
$S+$ & $L$ & $S-$ & $L$ & $L/S$ & $S-$ & $+1$ & rejected & rejected & rejected \\
$S-$ & $L$ & $S+$ & $L$ & $L/S$ & $S-$ & $+1$ & rejected & rejected & rejected \\
$S-$ & $L$ & $S-$ & $L$ & $L$ & $S+$ & $+1$ & rejected & rejected & rejected \\
$S+$ & $L$ & $L$ & $S+$ & $S+$ & $L/S$ & rejected & $+1$ & rejected & rejected \\
$S+$ & $L$ & $L$ & $S-$ & $S-$ & $L/S$ & rejected & $+1$ & rejected & rejected \\
$S-$ & $L$ & $L$ & $S+$ & $S-$ & $L/S$ & rejected & $+1$ & rejected & rejected \\
$S-$ & $L$ & $L$ & $S-$ & $S+$ & $L/S$ & rejected & $+1$ & rejected & rejected \\
$L$ & $S+$ & $S+$ & $L$ & $S+$ & $L/S$ & rejected & rejected & $+1$ & rejected \\
$L$ & $S+$ & $S-$ & $L$ & $S-$ & $L/S$ & rejected & rejected & $+1$ & rejected \\
$L$ & $S-$ & $S+$ & $L$ & $S-$ & $L/S$ & rejected & rejected & $+1$ & rejected \\
$L$ & $S-$ & $S-$ & $L$ & $S+$ & $L/S$ & rejected & rejected & $+1$ & rejected \\
$L$ & $S+$ & $L$ & $S+$ & $L/S$ & $S-$ & rejected & rejected & rejected & $-1$ \\
$L$ & $S+$ & $L$ & $S-$ & $L/S$ & $S+$ & rejected & rejected & rejected & $-1$ \\
$L$ & $S-$ & $L$ & $S+$ & $L/S$ & $S+$ & rejected & rejected & rejected & $-1$ \\
$L$ & $S-$ & $L$ & $S-$ & $L/S$ & $S-$ & rejected & rejected & rejected & $-1$ \\
\end{tabular}}
\end{ruledtabular}
\end{table*}


\section{\label{Sec2}Proposed test of Mermin inequality with three-qubit GHZ energy-time states}


The setups in in Refs.~\cite{shih93pla, pitt96pro} cannot
exclude local HV models like the one introduced in the previous
Section. Then, a natural question is how to exclude them and
perform a genuine test the Mermin inequality with energy-time
entanglement. In this Section we provide a solution based on a
new configuration which is a natural extension to three or more
parties of the scheme introduced in Ref.~\cite{cabe09prl}. The
advantage over the set up in Fig.~\ref{Fig1} discussed in
Sec.~\ref{Sec1} is that, in the case of perfect detectors, with
the new configuration the expected results cannot be simulated
with any local HV model.


\begin{figure}[h]
\begin{center}
\includegraphics[width=\columnwidth]{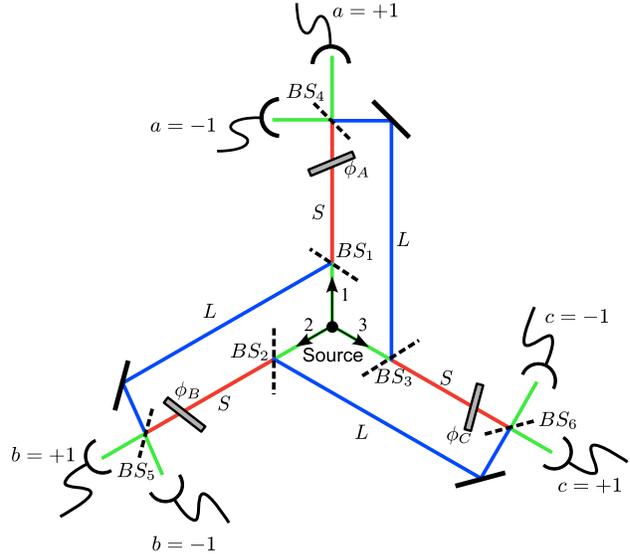}
\caption{Setup for preparing a three-qubit energy-time GHZ state.
All the beam splitters (BSs) are 50/50 BSs.}
\label{Fig2}
\end{center}
\end{figure}


The crucial difference between the setups of Figs.~\ref{Fig1}
and \ref{Fig2} is that while the geometry of the set up in
Fig.~\ref{Fig1} does not prevent that the selection and
rejection of events can be affected by the local phase
settings, the geometry of the setup in Fig.~\ref{Fig2},
prevents this possibility. Therefore, while local HV models for
experiments using the setup of Fig.~\ref{Fig1}, the decision of
being detected or not can depend on the local setting, in any
local HV model for experiments using the setup of
Fig.~\ref{Fig2}, the fact that the photon is detected or not
must be independent of the local phase settings; and there are
no such local HV models reproducing the quantum predictions.

To illustrate this difference, first consider a selected event:
the three photons have been detected at time $t_0$ (or at time
$t_1$). Although the phase setting of $\phi_A$, $\phi_B$, and
$\phi_C$ are, respectively, in the backward light cones of the
photons detected in Alice, Bob, and Carol's sides, as in the
setup of Fig.~\ref{Fig1}, the key point is that, in
Fig.~\ref{Fig2}, different values of the phase settings cannot
cause a selected event to become a rejected event, since this
would require a mechanism to make one detection to ``wait''
until the information about the setting in other side comes.
However, when this information has finally arrived, the phase
settings (both of them) have changed, so this information is
useless to base a decision on it.

On the contrary, for the setup of Fig.~\ref{Fig2}, there is no
physical mechanism preserving locality which can turn a
selected (rejected) event into a rejected (selected) event. The
selected events are independent of the local phase settings.
For the selected events, only the $+1/-1$ decision can depend
on the phase settings. This is exactly the assumption under
which the Mermin inequality (\ref{Mermin}) and their
generalizations to $n>3$ parties are valid. Therefore, an
experimental violation of (\ref{Mermin}) using the setup of
Fig.~\ref{Fig2} and the postselection procedure described
before provides a conclusive (assuming perfect detectors) test
of local realism using energy-time (or time-bin) entanglement.


\section{\label{Sec3}Generation of $N$-qu$N$it energy-time entangled states}


\subsection{Three-qutrit energy-time entangled states}

\begin{figure}[h]
\begin{center}
\includegraphics[width=\columnwidth]{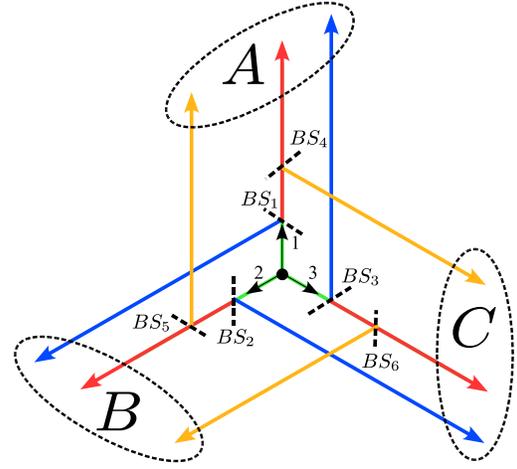}
\caption{Setup for preparing the state (\ref{state111222333}).
BS$_1$, BS$_2$, BS$_3$ have reflection coefficient $R=\frac13$, while
BS$_4$, BS$_5$, BS$_6$ have $R=\frac12$.}
\label{Fig3}
\end{center}
\end{figure}

An interesting feature of the setup in Fig.~\ref{Fig2} is that
it can be extended to prepare $n$-qu$n$it energy-time entangled
states whith potential applications in quantum information
processing. For instance, using the setup shown in
Fig.~\ref{Fig3}, we can prepare the three-qutrit state
 \beq
 \ket{\Psi}=\frac{1}{\sqrt3}(\ket{111}+\ket{222}+\ket{333}),
 \label{state111222333}
 \eeq
by using a source emitting three photons simultaneously at an
unknown time, and then postselecting the threefold
coincidences. The geometry of the setup is suitable for
three-qutrit Bell tests (i.e., is free of the problems
discussed in Sec.~\ref{Sec2}).


\begin{figure}[h]
\begin{center}
\includegraphics[width=\columnwidth]{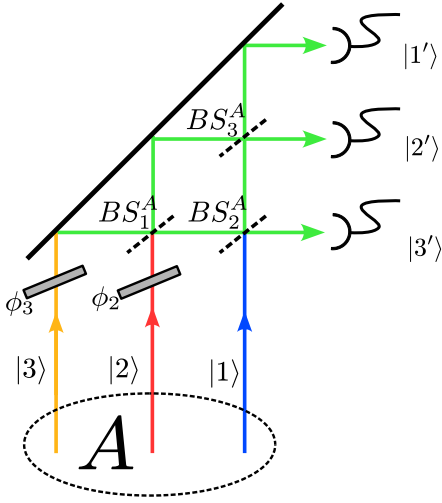}
\caption{Setup for the measurement of a qutrit state. The reflection coefficients are
given in \eqref{BSsFig4a}--\eqref{BSsFig4c}.}
\label{Fig4}
\end{center}
\end{figure}

The setup for performing one observer's local measurements is
shown in Fig.~\ref{Fig4}. The three BSs in Fig. \ref{Fig4},
written in the basis $\ket1,\ket2,\ket3$, are given by
 \begin{subequations}
 \begin{align}
\text{BS}^A_1 & =
\begin{pmatrix}
1 & 0 & 0 \\
0 & \frac{1}{\sqrt2} & \frac{e^{i\alpha}}{\sqrt2} \\
0 & \frac{1}{\sqrt2} & -\frac{e^{i\alpha}}{\sqrt2}
\end{pmatrix}, \label{BSsFig4a} \\
\text{BS}^A_2 & =
\begin{pmatrix}
\frac{\sqrt2}{\sqrt3} & 0 &\frac{e^{i\beta}}{\sqrt3} \\
0 & 1 & 0\\
\frac{1}{\sqrt3} & 0 &-\frac{\sqrt2 e^{i\beta}}{\sqrt3}
\end{pmatrix}, \\
 \text{BS}^A_3 & =
\begin{pmatrix}
\frac{1}{\sqrt2} & \frac{e^{i\gamma}}{\sqrt2} &0 \\
\frac{1}{\sqrt2} & -\frac{e^{i\gamma}}{\sqrt2} &0 \\
0 & 0 & 1
\end{pmatrix}. \label{BSsFig4c}
 \end{align}
 \end{subequations}
Therefore, $\text{BS}^A_1$ and $\text{BS}^A_3$ are 50/50 BSs,
while $\text{BS}^A_2$ has a reflection coefficient $R=\frac13$.
The action of the three BSs in Fig.~\ref{Fig4} corresponds to
the following unitary operator:
 \beq
\begin{aligned}
M&:=\text{BS}^A_3\text{BS}^A_2\text{BS}^A_1\\
&=
\frac{1}{\sqrt3}
\begin{pmatrix}
1 & \frac{\sqrt3e^{i\gamma} +e^{i\beta}}{2}& \frac12e^{i\alpha}(\sqrt3e^{i\gamma} -e^{i\beta}) \\
1 & \frac{-\sqrt3e^{i\gamma} +e^{i\beta}}{2}& -\frac12e^{i\alpha}(\sqrt3e^{i\gamma} +e^{i\beta})\\
1 & -e^{i\beta}& e^{i(\beta+\alpha)}
\end{pmatrix}.
\end{aligned}
 \eeq
By choosing $\beta=\frac\pi3$, $\gamma=-\frac\pi6$
and $\alpha=\pi/3$, we obtain
 \beq
M= \frac{1}{\sqrt3}
\begin{pmatrix}
1 & 1 & 1\\
1 & e^{i\frac{2\pi}3} & e^{i\frac{4\pi}3}\\
1 & e^{i\frac{4\pi}3} & e^{i\frac{8\pi}3}
\end{pmatrix}.
 \eeq
By inserting the three phases $\phi_i$, we obtain
 \beq M=
\frac{1}{\sqrt3}
\begin{pmatrix}
1 & e^{-i\phi_2} & e^{-i\phi_3}\\
1 & e^{i\frac{2\pi}3}e^{-i\phi_2} & e^{i\frac{4\pi}3}e^{-i\phi_3}\\
1 & e^{i\frac{4\pi}3}e^{-i\phi_2} & e^{i\frac{2\pi}3}e^{-i\phi_3}
\end{pmatrix}.
 \eeq
This measurement projects onto the basis
 \beq
\ket{1'}=M^\dag\ket1\,,\quad \ket{2'}=M^\dag\ket2\,,\quad
\ket{3'}=M^\dag\ket3\,,\quad
 \eeq
 given by
 \begin{subequations}
 \begin{align}
 \ket{1'}&=\frac{1}{\sqrt3}(\ket1+e^{i\phi_2}\ket2+e^{i\phi_3}\ket3),\\
 \ket{2'}&=\frac{1}{\sqrt3}(\ket1+e^{i(\phi_2-\frac{2\pi}3)}\ket2+e^{i(\phi_3-\frac{4\pi}3)}\ket3),\\
 \ket{3'}&=\frac{1}{\sqrt3}(\ket1+e^{i(\phi_2-\frac{4\pi}3)}\ket2+e^{i(\phi_3-\frac{2\pi}3)}\ket3).
 \end{align}
 \end{subequations}


\subsection{Generalization to $n$ qu$n$its}

\begin{figure}[h]
\begin{center}
\includegraphics[width=0.95\columnwidth]{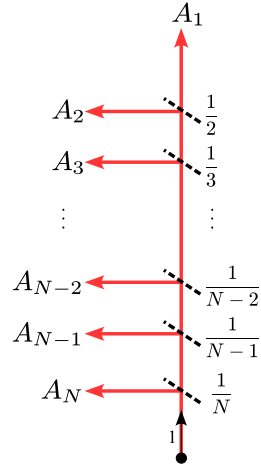}
\caption{Generation of $n$-particle qu$n$it state. The reflection coefficients are shown
on the right side of the corresponding BS.}
\label{Fig5}
\end{center}
\end{figure}

Interestingly, the setup can be extended to prepare $n$-qu$n$it
energy-time entangled states with $n >3$. For each particle we
use a scheme given in Fig.~\ref{Fig5} to generate a qu$n$it.
Each mode is sent to a different party $A_i$. Then, by using a
scheme similar to that proposed in \cite{reck94prl} we can
measure the qu$n$it.


\begin{figure}[h]
\begin{center}
\includegraphics[width=0.95\columnwidth]{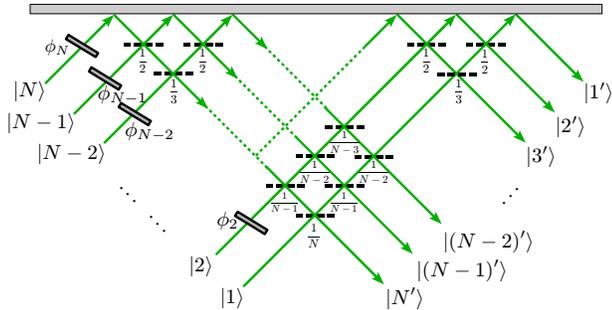}
\caption{Measurement of a qu$n$it state. The reflection coefficients are shown
below the corresponding BS.}
\label{Fig6}
\end{center}
\end{figure}

The BSs described in Fig.~\ref{Fig6} can be set to produce the following
unitary transformation
 \beq
 \mathcal U=\frac1{\sqrt N}
\begin{pmatrix}
1 & 1 & 1 & \cdots & 1 \\
1 & \omega& \omega^2&\cdots & \omega^{N-1} \\
1 & \omega^2& \omega^4&\cdots & \omega^{2(N-1)} \\
\vdots &&&&\vdots \\
1 & \omega^{N-1}& \omega^{2(N-1)}&\cdots & \omega^{(N-1)^2} \\
\end{pmatrix},
 \eeq
where
 \beq
 \omega=e^{\frac{2\pi i}{N}}.
 \eeq
We have
 \beq
 \mathcal U_{ij}=\omega^{(i-1)(j-1)}.
 \eeq
With the phase shift we measure on the following basis:
 \begin{subequations}
 \begin{align}
 \ket{1'}&=\frac{1}{\sqrt N}(\ket1+e^{i\phi_2}\ket2+\cdots +e^{i\phi_N}\ket N),\\
 \ket{2'}&=\frac{1}{\sqrt N}(\ket1+\bar\omega e^{i\phi_2}\ket2+\cdots +\bar\omega^{N-1}e^{i\phi_N}\ket N),\\
 \ket{3'}&=\frac{1}{\sqrt N}(\ket1+\bar\omega^2 e^{i\phi_2}\ket2+\cdots +\bar\omega^{2(N-1)}e^{i\phi_N}\ket N),\ldots,\\
 \ket{N'}&=\frac{1}{\sqrt N}(\ket1+\bar\omega^{N-1} e^{i\phi_2}\ket2+\cdots +\bar\omega^{(N-1)^2}e^{i\phi_N}\ket
 N).
 \end{align}
 \end{subequations}


\section{\label{Sec4}Sources with unknown emission time}


So far, we have assumed that we have sources capable to emit
three or more particles at the same unknown time. However, to
our knowledge, no such sources exist. This forces us to use, in
actual experiments, sources in which pairs of particles are
emitted at different unknown times. The use of these sources
does not solve the problem described in Sec.~\ref{Sec2}, just
makes the problem more complex to analyze. The conclusion is
still the same: Franson-like Bell experiments admit local HV
models reproducing the quantum predictions, even when we use
these sources. The aim of this Section is to show that these
sources can be used with the schemes introduced in
Sec.~\ref{Sec3}, and still local HV models reproducing the
quantum predictions are impossible.

\begin{figure}[t]
\begin{center}
\includegraphics[width=1\columnwidth]{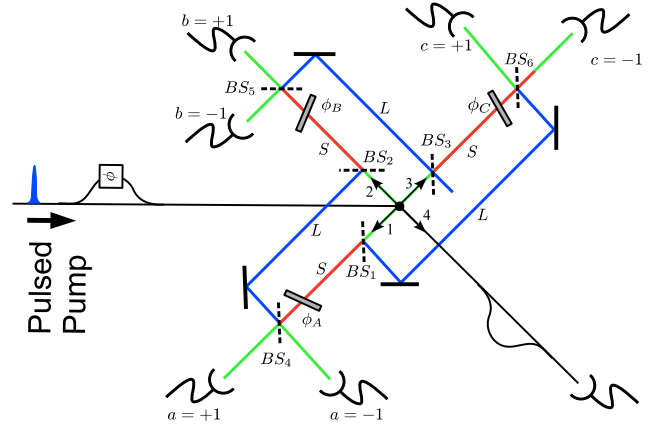}
\caption{Realistic setup for a test of the three-party Mermin inequality
with energy-time entangled photons.}
\label{Fig7}
\end{center}
\end{figure}

For instance, in order to test the Mermin inequality on
three-photon GHZ state we would need a (nonexistent) source
emitting three photons at the same unknown time. However, a
feasible realization is the one illustrated in Fig.~\ref{Fig7}.
A femtosecond pulsed laser (with very low coherence time) is
injected into a Mach-Zender (MZ) interferometer before shining
the nonlinear crystal, from which two independent pairs are
emitted at different times. The rest of the setup is similar to
the one described in Sec. \ref{Sec3}.


If $t_0$ ($t_1=t_0+\delta t$) are the arrival time of the short
(long) arm pump pulse, where $\delta t$ is the path difference
and if the two photon pairs are $(1, 2)$ and $(3, 4)$, then,
generated state is given by
 \beq
 \begin{aligned}
 \ket{\psi}=&\frac{1}{2}(\ket{t_0}_1\ket{t_0}_2\ket{t_0}_3\ket{t_0}_4+
 \ket{t_1}_1\ket{t_1}_2\ket{t_1}_3\ket{t_1}_4\\
 &+\ket{t_0}_1\ket{t_0}_2\ket{t_1}_3\ket{t_1}_4+
 \ket{t_1}_1\ket{t_1}_2\ket{t_0}_3\ket{t_0}_4).
 \end{aligned}
 \label{state}
 \eeq
Note that, if the four photons could be generated at the same
time, we would have only the first two terms in \eqref{state}.
The latter two terms contribute to events that are detected on
different sides and are not coincident. In order to discard
them, we should shorten the coincidence windows.

Now the question is whether the selected/rejected events could
have been rejected/selected events for a different value of one
of the local settings. The key point to see that this cannot
happen is to remember that photons 1 and 2 (3 and 4), when the four photons
arrive at four different locations, are always detected at the same time. 
Then, only a nonlocal mechanism can
change the arrival time of {\em both} photons due to a
different local setting in {\em one} of the photons. In
principle, the detection of photon 1 at time $t_0$ or $t_1$
could depend of the local setting. The problem for any local HV
model reproducing the quantum predictions is that photon 2
should be detected at the same time, and this requires nonlocal
communication. Therefore, the use of these sources do not cause
any fundamental problem if the detectors are perfect.


\section{\label{Sec5}Conclusions}


Franson's energy-time entanglement was a great achievement
because provided a new experimentally feasible method to
generate photonic entanglement. However, the fact that (without
supplementary assumptions), the outcomes of actual Bell-CHSH
experiments, and even those of ideal experiments, can be
reproduced with local HV models weakens the power of the idea.
In a previous paper \cite{cabe09prl}, we proposed a way to
solve this problem which has been implemented in actual
experiments. Then, a natural question is whether the same
problem affects previously proposed extensions of Franson's
setup to the multipartite case. It does. In this paper, we have
shown how to extend our previous proposal to fix the problem in
the multipartite case and discussed possible applications of
this extension. Specifically, we have shown, that, in
principle, there is no fundamental obstacle to perform
experimental tests of the $n$-party Bell inequality ($n \ge 3$)
proposed by Mermin \cite{merm90prl} with energy-time entangled
photons prepared in $n$-qubit GHZ states, and how to produce a
class of $n$-particle $n$-level entangled states with potential
applications in quantum information processing, using
energy-time.


\section{Acknowledgments}


We thank J. Franson, J.-\AA. Larsson, and M. \.{Z}ukowski for
useful conversations. This work was supported by the Spanish
Ministry of Science and Innovation's project No. FIS2008-05596,
from the Junta de Andaluc\'{\i}a's Excellence Project No.
P06-FQM-02243, and by Finanziamento Ateneo 07 Sapienza
Universit\'{a} di Roma.





\end{document}